\documentstyle{article}
\def\be{\begin{equation}}
\def\ee{\end{equation}}
\def\bea{\begin{eqnarray}}
\def\eea{\end{eqnarray}}
\def\la{\label}
\def\ci{\cite}
\def\lr{\left}
\def\rr{\right}

\begin{document}

\begin{flushright}
OUTP-93-33P \\
\end{flushright}
\vspace{15mm}

 \begin{center}
 {\large\bf Unification Scale in String Theory}\\
 \vspace{7mm}
{\large  A. de la Macorra}\footnote{From 1 January 1994 E-mail:
macorra@teorica0.ifisicacu.unam.mx}\\ [5mm]
{\em Department of Physics, University of Oxford,\\
  1 Keble Rd, Oxford OX1 3NP}\\ [8mm]
 \end{center}
\vspace{2mm}
\begin{abstract}
\noindent
We study the unification scale and gauge coupling constant in
 4D string theory.  We show that the fine structure constant is
 determined by the dimension of the hidden gauge group and only
 $SU(6)$ and $SO(9)$ are consistent with minimal string unification
while  the unification scale can be of order of $10^{16}\,GeV$.
\end{abstract}

\renewcommand{\thefootnote}{\arabic{footnote})}
\setcounter{footnote}{0}

\newpage

	String theory provides with the only possibility, up till now, of unifying all
interactions \ci{r1}.
As formulated in the critical dimensions it has only one free parameter which
is
taken as the Planck mass. But once it is compactified to four dimensions much
of its uniqueness is lost because there are a great number of consistent string
vacua \ci{r25}-\ci{r36}. Nevertheless it is possible to study general 4D string
models by concentrating
on  properties shared by all models and consider the model dependent
quantities as free parameters.  These models contain, after compactification,
 two separate gauge
groups $E_{6}\times E_{8}'$ or subgroups there of. The group $E_{6}$,
called the visible sector, is
supposed to have the standard model as a subgroup  while the other group
$E_{8}'$ (or a subgroup) is referred to as the hidden gauge
 group.  In many cases these two sectors interact through gravity
only.  In this context supersymmetry (SUSY)  will be broken
in the hidden sector via a gaugino condensate
and it will be transmitted to the observable sector  through gravity.
The scale at which the condensate is  formed is related to
the  condensation scale,  defined as the scale where the
gauge coupling constant becomes strong. Clearly it will vary for
different gauge groups.
  Since in string theory  all  gauge coupling
constants are unified at the string scale and  are mainly given by
the v.e.v. of the dilaton field $S$, the condensation scale for each gauge
group will be determined by its one-loop beta function coefficient $\beta_{0}$.
The larger $\beta_{0}$ is, the larger  the condensation scale will be, and we
will then expect that SUSY is broken by the gauge group with largest
$\beta_{0}$. Clearly, some relevant model dependent quantities are then the
hidden gauge group and its spectrum which determine the scale of SUSY breaking.
The standard model can then be viewed as a globally
supersymmetric model with explicit soft supersymmetric breaking terms and the
study of these
terms is relevant in determining the viability of  the models.

Another generic property is the invariance of the effective
Lagrangian under duality symmetry \ci{r7}. Although it has only been
proved to be exact to all orders in perturbation theory \ci{r38} one
expects that non-perturbative effects will respect duality. This
symmetry is intimately related to the contribution of the
infinite number of Kaluza-Klein modes, always present in string
theory. These modes become relevant in the low energy effective
theory for small values of
the compactified radius given by the real part of the (1,1)
moduli.  In the simplest case, the moduli fields transform
under duality as an element of the SL(2,Z) group \ci{r5,r34} and in this
paper we will only consider those moduli.

In string theory the gauge coupling constant and the unification scale are
determined dynamically and are given in terms of the v.e.v.s of the dilaton and
moduli fields.  Assuming
that the low energy model is that of the minimal supersymmetric standard model
(MSSM)
unification of the gauge coupling constants of the standard model restricts
the values of the gauge coupling constant at the unification scale and the
unification scale
to a narrow band \ci{r58}. In particular the
unification scale should be of order  $10^{16} GeV$ but in string theory it
is usually much larger.  The reason is that  the unification scale is given in
terms of a modular invariant  function of the moduli fields.  If
the v.e.v.s of the moduli  is of order one, which is their v.e.v. at "tree"
level\footnote{By tree level  we understand that it does not include loop
corrections of the strongly binding effects} \ci{r49,r50},  the unification
scale is
of order of the string scale. However, it was recently shown that once loop
corrections of the strong binding effects, which lead to a gaugino condensate,
are taking into account,  the v.e.v. of the
moduli can be much larger \ci{r52,r56}.  Furthermore,
 their v.e.v.s may differ allowing for  squeezed orbifolds which are better
candidates for  minimal string unification \ci{r34}.
Here we will study whether it is possible to obtain the desired values for the
unification scale and gauge coupling constant.

	We study the possibility of having unification
of the gauge coupling constants assuming the minimal string
unification scheme (MSU). This scheme consists of having the chiral matter
content in the visible sector of the theory  to be just that of the
minimal supersymmetric model (MSSM) while the hidden sector remains
unspecified.   Recent precision measurements for
the weak coupling and strong coupling constant at LEP have permitted
to refine the analysis of the unification of the gauge coupling
constants. The $SU(5)$ prediction is now ruled out but the
supersymmetric extension is in good agreement with the experimental
values \ci{r58}.

The gauge coupling constants in the standard model (SM) are measured
at the electroweak energy scale. Using the well-known
renormalization group equation these couplings can be  determined
 for any other  energy scale.
The gauge coupling constants have a logarithmic evolution
with respect to  energy and  at
the one-loop level they are given by
\be
g^{-2}_{a}(\Lambda)=k_{a}g^{-2}_{0}(\Lambda_{0})+2b_{a}ln(\frac{\Lambda}{\Lambda_{0}}).
\la{f1}\ee
 $\Lambda_{0}$ is the energy scale at which the gauge coupling
constant $g_{0}$ is measured, $g_{a}$ is the gauge coupling constant
at the  arbitrary scale  $\Lambda$, and $b_{a}$ is the one-loop $\beta$
function coefficient for the $G_{a}$ gauge group while $k_{a}$ is its
corresponding Kac-Moody level. In the SM the beta function coefficients
are given by $16\pi^{2}b_{3}=7, 16\pi^{2}b_{2}=-19/6$ and
$16\pi^{2}b_{1}=-41/6$ with
$k_{3}=k_{2}=k_{1}3/5=1$
for $SU(3), SU(2)$ and $U(1)$  respectively.

As mentioned above, it has recently been shown that the evolution of the gauge
coupling constants in the SM do not become unified, i.e. they do not
meet for any given energy. However, interestingly enough they do meet if one
includes supersymmetry (SUSY) and assumes the  minimal supersymmetric
standard model (MSSM).The
main reason is that in the latter case the one-loop beta function
coefficients change due to the inclusion of the extra states and they
become $16\pi^{2}b_{3}=3, 16\pi^{2}b_{2}=-1$ and $16\pi^{2}b_{1}=-11$
 and the strong coupling constant
becomes less asymptotically free. For these values of the one-loop
beta function coefficient
and assuming that the supersymmetric breaking scale $M_{ss}$ is of order
$10^{3\pm 1}\, GeV$,
the gauge coupling constants are unified at a value for the fine
structure constant given by (in the minimal subtraction scheme
$\overline{MS}$)  \ci{r58}
\be
\hat{\alpha}^{-1}_{gut}\equiv \lr(\frac{g^{2}_{gut}}{4\pi}\rr)^{-1}=25.7\pm 1.7
\la{f2}\ee
and unification scale
\be
\Lambda_{gut}=10^{16\pm0.3}\,GeV.
\la{f3}\ee
Though in the MSSM one has three free parameters $M_{ss}, g_{gut}$
and $\Lambda_{gut}$ to predict three coupling constants, meaning that
 there will always be a solution to the unification of the
gauge coupling constant, this fact  does not mean that the unification
scheme is empty because the values of the
parameters  $M_{ss}, g_{gut}$ and $\Lambda_{gut}$ are very much restricted
by phenomenological constraints. In fact, one requires
$M_{Planck}>\Lambda_{gut}>10^{15}\,GeV$  to avoid a fast
decaying proton, $M_{ss}$
must be at the most of order 1 TeV to explain the mass hierarchy
problem and larger then $100\, GeV$ to prevent light SUSY states that
would have been detected already. Finally
$g_{gut}^{2}$ must be positive and small so that we stay in the
perturbative regime.

The running of the couplings in string  theory
is given by \ci{r17}-\ci{r18}
\be
\frac{1}{g_{a}^{2}(\Lambda)}=\frac{k_{a}}{g^{2}_{s}}+b_{a}ln(\frac{\Lambda^{2}}{M_{a}^{2}})+\Delta_{a}
\la{f17}\ee
where $b_{a}=\frac{1}{16\pi^{2}}(3C(G_{a})-\Sigma_{R_{a}}h_{R_{a}}T(R_{a}))$ is
the N=1 $\beta$-function coefficient , $C(G_{a})$ the quadratic Casimir
operator and
$h_{R_{a}}$ the number of
chiral fields in a representation $R_{a}$ (the latin indices of the
beginning of the alphabet ($a,b,..$) represent gauge indices while those
in the middle of the alphabet ($i,j,...$) refer to the type of moduli
$T_{i}$).   $M_{a}$, as defined below, is the
renormalization scale  below
which the coupling constant begins to run. The $o$-index refers
to the hidden sector gauge group only
while the indices $a$ and $b$
correspond to  generic gauge groups (visible or hidden sector).

The existence of an infinite number
of massive states (Kaluza and winding states), above the string scale,
give rise to string threshold contributions $\Delta_{a}$
which are relevant to the determination of the coupling constant at
the string scale.  These threshold effects can be directly calculated
by computing
world-sheet string amplitudes  involving external gauge fields and
moduli  \ci{r17a,r18}, and has been done for (2,2) symmetric orbifold
compactification.
Another possible way to calculate the threshold corrections is by
imposing target space modular  invariance and the  cancellation of
target space modular anomalies.

The scale $M_{a}$, below which the coupling constant starts to run,
is in general a moduli dependent quantity
\be
M^{2}_{a}=\Sigma_{i}\,(T_{ri})^{\alpha_{i}}\,M_{s}^{2}
\la{f18}\ee
where $T_{ri}=(T+\bar{T})_{i}$ and
the constant $\alpha_{i}$ is model and gauge dependent and $M_{s}$ is
the string scale \footnote{Numerically the string scale is
$0.7\,g_{gut}\,10^{18}  GeV \ci{r18}$}.
In the case of a single  overall (1,1) moduli $T_{i}=T$, $i=1,2,3$,
with $\alpha=\Sigma\alpha_{i}=-1$ one obtains the ``naive''
field theoretical expression
\[
M_{a}=(Re S\, Re T)^{-1/2}.
\]
The contribution to the gauge coupling constant from the moduli fields
in eq.(\ref{f18}) can be calculated from the
anomalous triangle diagrams with two gauge bosons and several moduli
fields as external legs and massless gauginos and charged (fermionic)
matter fields circulating inside the loops.

The threshold term  in eq.(\ref{f17}) is given by
\be
\Delta_{a}=\Sigma_{i}(b'^{i}_{a}-k_{a}\delta^{i}_{GS})ln|\eta(T_{i})|^{4},
\la{f19}\ee
\[
b'^{i}_{a}=\frac{1}{16\pi^{2}}(C(G_{a})-\Sigma_{R_{a}}h_{R_{a}}T(R_{a})(1+2n^{i}_{R_{a}}))
\]
where $n^{i}$ is the modular weight for a chiral matter superfield
with respect to the $i$-moduli and
$\eta(T)$ the Dedekind-eta function.
In the case of an overall moduli
$b'_{a}=\Sigma_{i}b'^{i}_{a}=\frac{1}{16\pi^{2}}(3C(G_{a})-\Sigma_{R_{a}}h_{R_{a}}T(R_{a})(3+2n_{R_{a}}))=b_{a}-2\Sigma_{R_{a}}h_{R_{a}}T(R_{a})(1+n_{R_{a}})$
with $n_{R_{a}}=\Sigma_{i}n^{i}_{R_{a}}$.

The universal Green-Schwarz coefficient $\delta_{GS}$ in eq.(\ref{f19}) is
needed to cancel, using  the Green-Schwarz mechanism, the gauge
independent part of the target space modular anomaly. The threshold
contribution of the massive fields, $b'^{i}_{a}-k_{a}\delta^{i}_{GS}$,
is in general non-vanishing if at least one of the orbifold twists
leaves the $i$-plane unrotated. In this sector, the massive spectrum
is N=2 space-time supersymmetric and $b^{i}_{a}-k_{a}\delta^{i}_{GS}$
is proportional to the  N=2
$\beta$-function coefficient which is in general non zero. On the other
hand if all orbifold twists rotate a specific plane, than the spectrum
is N=4 supersymmetric and $b'^{i}_{a}-k_{a}\delta^{i}_{GS}=0$, giving no
threshold contribution. In this case the gauge coupling constant is
independent of the $T_{i}$ moduli and one has that $b'^{i}_{a}/k_{a}$
must be equal for all gauge groups, i.e.  $b'^{i}_{a}/k_{a}=b'^{i}_{b}/k_{b}$.

Eq.(\ref{f17}) can be rewritten in a similar form as eq.(\ref{f1})
\be
g^{-2}_{a}(\Lambda)=k_{a}g^{-2}_{gut}(\Lambda^{a}_{gut})+2b_{a}ln(\frac{\Lambda}{\Lambda^{a}_{gut}})
\la{f20}\ee
with the unification scale defined by
\be
\Lambda^{a}_{gut}=M_{s}(\Pi_{i}T_{ri}|\eta(T_{ri})|^{4})^{\alpha^{i}_{a}/2}
\la{f21}\ee
where
\be
\alpha_{a}^{i}\equiv \frac{\delta^{i}_{GS}k_{a}-b'^{i}_{a}}{b_{a}}
\la{f13}\ee
and the gauge coupling constant at the unification scale given by
\be
g^{-2}_{gut}=\frac{Y}{2}
\ee
where at tree level  $Y$ is given in terms of the v.e.v. of the dilaton field
$S$.
For two gauge coupling constants to become equal, i.e.
$\frac{g^{2}_{a}(\Lambda_{gut})}{k_{a}}=\frac{g^{2}_{b}(\Lambda_{gut})}{k_{b}}=g^{2}_{gut}(\Lambda_{gut})$,
at the unification scale $\Lambda_{gut}$   the coefficients defined
in eq.(\ref{f13}) for different gauge groups must  be
the same, i.e. $\alpha_{a}^{i}=\alpha_{b}^{i}$.
{}From eq.(\ref{f20}) one obtains the condensations scale, defined as the scale
where the gauge coupling constant becomes strong, and it is given by
\be
\Lambda_{c}^{a}=\Lambda_{gut}^{a}\, e^{-k_{a}Y/4b_{a}}.
\ee
{}From eq.(\ref{f21}) we note that a
unification scale smaller then the string scale  necessarily requires
the exponent to be positive, $\alpha>0$ since the modular
invariant function $T_{ri}|\eta_{i}|^{4}< 1$.
One can eliminate the Green-Schwarz term
in eq.(\ref{f21}) and the unification scale becomes \ci{r54}
\be
\Lambda_{gut}=
M_{s}(\Pi_{i}T_{ri}|\eta(T_{ri})|^{4})^{\frac{b'^{i}_{a}-b'^{i}_{b}}{2(b_{b}-b_{a})}}
\la{f23}\ee
and it is completely specified once the v.e.v. of the moduli and the
$b_{0}$ and $b'_{i}$ coefficients are determined. A positive
$\alpha$ or equivalently a positive exponent in eq.(\ref{f23})
forces the compactified space to have chiral matter fields with
modular weights different then -1/3 (untwisted fields have modular
weights -1/3) otherwise $b'^{i}_{a}=b_{a}/3$.

	In string theory the different parameters of the low energy
theory like unification scale, supersymmetry breaking scale and gauge
coupling constant at the unification scale are dynamically determined once
 supersymmetry  is broken.
The most common and probably the best way for breaking SUSY is via
gaugino condensate \ci{r9}.  In order to study the breaking of SUSY
one determines the effective potential and   the  vacuum.
The effective interaction involving the
gaugino bilinear can be obtained
by demanding the complete Lagrangian
 to be anomaly free under the R-symmetry  under which the
gauginos transform non-trivially.
The low energy degrees of freedom are then
the dilaton field $S$, moduli fields $T_{i}$, chiral
matter fields $\varphi_{i}$, gauge fields
and the Goldstone
mode $\Phi$ associated with the spontaneously broken   R-symmetry plus
their supersymmetric partners. In orbifold compactification there are
always three diagonal moduli
whose real parts represent the size of the
compactified complex plane and here we will only consider these moduli.
The 4D string model is given by an N=1 supergravity theory and it is specified
once the Kahler potential $G=K+ln\frac{1}{4}|W|^{2}$ and the gauge kinetic
function $f$ are given.
The Kahler potential, superpotential and gauge kinetic functions are
\ci{r37}-\ci{r51}
\be
K=-ln\lr(S+\bar{S}+2\Sigma_{i}(k_{0}\delta_{GS}^{i}-b'^{i}_{0})lnT_{ri}\rr)-\Sigma_{i}ln(T_{ri})+K^{i}_{i}|\varphi_{i}|^{2},
\la{d38}\ee
\be
W_{0}=\Pi_{i}\eta^{-2}(T_{i})\,\,\Phi\,+\,W_{m}
\la{d39}\ee
and \ci{r52}
\be
f=f_{0}+\frac{2}{3}b_{0}ln(\Phi)
\la{d40}\ee
respectively.  $W_{m}$ is the superpotential for the chiral matter
superfields and the gauge kinetic function at the string scale is
given by \ci{r17}-\ci{r18},\ci{r39}
\be
f_{0}=S+2\Sigma_{i}(b'^{i}_{a}-k_{a}\delta^{i}_{GS})ln\,[\eta(T_{i})^{2}]
\la{d41}\ee
where $\eta$ is the Dedekind-eta function and $b_{0}$
the one-loop beta function for the hidden gauge group. The coefficient
$\delta_{GS}$ is the Green-Schwarz term needed to cancel the gauge
independent part of the target modular anomaly and $b'^{i}_{0}$ define in
eq.(\ref{f19}).

Through the equation of
motion of the auxiliary field of $\Phi$, the scalar component is given in terms
of the gaugino bilinear  of the hidden sector  \ci{r52}
\[
\phi=\frac{e^{-K/2}\xi}{2\Pi_{i}\eta^{-2}(T_{i})}\,
\bar{\lambda}_{R} \lambda_{L}
\]
with $\xi=2b_{0}/3$.
The model described in eqs.(\ref{d38}-\ref{d40}) is anomaly free and
duality invariant.  The duality transformation for the fields read
\bea
S&\rightarrow& S + 2\Sigma_{i}(k_{a}\delta^{i}_{GS}-b'^{i}_{a})ln(icT_{i}+d),
\nonumber\\
T_{i} &\rightarrow& \frac{a T_{i}-ib}{icT_{i} + d},
\la{d42}\\
\phi&\rightarrow& \phi,
\nonumber\eea
with $a,b,c,d\, \epsilon\, Z $ and $ad-bc=1$.
This model generates
a four-Gaugino interaction and  reproduces the tree
level scalar potential used by other parameterizations of  gaugino
condensate \ci{r50,r51,r55}. It  also  permits the   determination of the
radiative
corrections and use  of NJL technique  \ci{r10} to extract non-perturbative
information in the regime of strong coupling.
After minimizing the complete scalar potential (tree
level plus one-loop potential),  the vacuum structure
is quite different than the tree level one, thus  permitting us to
 find a stable
solution for the dilaton with the inclusion of a single  gaugino
condensate. We will  show that the value of the v.e.v.'s of the
dilaton and moduli fields can give a good prediction of the fine
structure constant  at the unification scale and unification scale
allowing for minimal string unification  to work.

In supergravity theory the tree level scalar potential is given by \ci{r15}
\be
V_{0}= h_{i}(G^{-1})^{i}_{j}h^{j}-3m^{2}_{3/2}
\la{f8}\ee
where the auxiliary fields are
$h_{i}=-e^{G/2}G_{i}+\frac{1}{4}f_{i}\bar{\lambda}_{R}\lambda_{L}$.
For the choice of Kahler potential and gauge kinetic function given in
eqs(\ref{d38}-\ref{d40}) one has
\be
V_{0}=e^{G}\,B_{0}=
\frac{1}{4}e^{K}\Pi_{i}|\eta(T_{i})|^{-4}\,\,|\phi|^{2}\,B_{0}
\la{d35}\ee
with
\be
B_{0}=
(1+\frac{Y}{\xi})^{2}+\Sigma_{i}\frac{Y}{Y+a_{i}}(1-\frac{a_{i}}{\xi})^{2}\,
\frac{T^{2}_{ri}}{4\pi^{2}}|\hat{G}_{2}(T_{i})|^{2}-3,
\la{d36}\ee
$a_{i}=2(k_{a}\delta_{GS}^{i}-b'^{i}_{a})$
and $Y=S+\bar{S} + 2\Sigma_{i}(k_{0}\delta_{GS}^{i}-b'^{i}_{0})lnT_{ri}$.
 The gravitino mass is given by
\be
m^{2}_{3/2}=\frac{1}{4}e^{K}\Pi_{i}|\eta(T_{i})|^{-4}\,\,|\phi|^{2}.
\la{f11}\ee
For a fixed value of $S$ the extremum solution to eq.(\ref{d35}) for the moduli
fields gives a
v.e.v. of $<T_{i}>\simeq 1.2$ \ci{r49,r50}.  This value is independent of $S$
and yields an unification scale of order of the string scale much larger than
the required value (cf. eq.(\ref{f3})).
Furthermore, there is
no stable solution in the dilaton
direction, it is a runaway potential for $S\rightarrow \infty$ and it
is unbounded from below for $S\rightarrow 0$. This is not surprising
because we haven't included the contribution from loop corrections of
the strong coupling
constant responsible for the gaugino binding. These contributions can be
 calculated using the Coleman-Weinberg one-loop potential
 $V_{1}$ and it is given by  \ci{r23,r57},
\be
V_{1}=\frac{1}{32\pi^{2}} Str \int d^{2} p\, p^{2}
ln(p^{2}+M^{2}) \la{a24}\ee
where $M^{2}$  represents the square mass matrices and $Str$ the
supertrace.
By solving the mass gap equation
$\frac{\partial}{\partial \phi}(V_{0}+V_{1})=0$ one is effectively
summing an infinite number of gaugino bubbles.  After integrating $\phi$ out
($\phi=\Lambda_{c}^{3}$) one can
minimize $V_{0}+V_{1}$ with respect to the moduli fields and
one obtains that the extremum equations are satisfied if  they either take the
dual invariant values ($<T>=1,e^{i\pi/6}$) or they take a common
``large'' value (i.e. $<T_{i}>=<T_{j}>$). For those moduli $T_{i}$ that take a
``large''
v.e.v. the corresponding $\alpha_{0}^{i}$ parameter defined in eq.(\ref{f13})
must be
the same, i.e. $\alpha_{0}^{i}=\alpha_{0}^{j}$, and
from now on  we will drop the $i$-index in $\alpha_{0}$.

The extremum equations for the dilaton and moduli  fields
yield the constraint  \ci{r56}
\be
B_{0}=\frac{9}{2\gamma
b_{0}^{2}}\lr(1+\frac{2\alpha_{0}-1}{3\alpha_0-1}\,\epsilon\rr)^{-1}
\la{e30}\ee
and
\be
\epsilon\equiv
x_{g}ln(x_{g})|_{min}=\frac{4b_{0}}{Y}(3\alpha_0-1)
\la{e31}\ee
where $B_{0}$ is given in eq.(\ref{d36}),
 $x_{g}=\frac{m^{2}_{g}}{\Lambda_{c}^{2}}$ with  $m^{2}_{g}=m^{2}_{3/2}
\frac{\xi^{2} B_{0}^{2}}{4 (Ref)^{2}}$ the
gaugino mass  square and $\Lambda_{c}$ the condensation
scale. Since at the minimum $x_{g}<1$
eq.(\ref{e31})  requires $\alpha_0<1/3$.
After eliminating $\phi$ the gravitino mass becomes
\be
m_{3/2}^{2}=\frac{1}{4Y}\,M^{6}_{s}\Pi_{i}(T_{ri}|\eta(T_{i})|^{4})^{3\alpha^{i}_{0}-1}\,e^{-3Y/2b_{0}}
\la{ff3}\ee
and
\be
x_{g}=(\frac{b_{0}}{6})^{2}
\frac{B^{2}_{0}}{Y}\Pi_{i}(T_{ri}|\eta(T_{i})|^{4})^{\alpha_{0}^{i}-1}e^{-Y/2b_{0}}.
\la{f16}\ee

{}From eqs.(\ref{e30}) and (\ref{e31}) it follows that for reasonable solutions
the dominant term
in $B_{0}$ is given by the contribution from v.e.v. of the auxiliary
field  of the dilaton $h_{s}$ and one  can approximate
$B_{0}\simeq (\frac{3Y}{2b_{0}})^{2}$. The v.e.v. of $Y$
and $T_{i}$ are then given in terms of the dimension of the hidden gauge
group, its  one-loop N=1 $\beta$-function coefficient and
$\alpha$ by
\be
Y\simeq
8\pi\sqrt{\frac{1}{n_{g}}\lr(1+\frac{2\alpha-1}{3\alpha-1}\,\epsilon\rr)^{-1}},
\la{e32}\ee
\be
\Pi_{i}T_{ri}|\eta(T_{i})|^{4}=\frac{\xi^{3}B_{0}^{3}m_{3/2}}{32\,Y\,x^{3/2}_{g}}.
\la{e33}\ee
Eq.(\ref{e33}) is obtained from eqs.(\ref{ff3}) and (\ref{f16})
and the sum
 is over all moduli that acquire a
v.e.v. different from the dual points.  To leading order the v.e.v of the
moduli is
\be
\Sigma_{i}(1-\alpha_{0i})T_{ri} =\frac{3Y}{\pi b_{0}}.
\ee
If the gauge group is broken down from $E_{8}$ to a lower rank group
such as  $SU(N)$ with $5\le N \le 9$ \ci{r56},
as can be easily done by compactifying on an orbifold with Wilson
 a large hierarchy  can be obtained with  only one gaugino
condensate.

In any given string vacuum the gauge groups and particle spectrum are
entirely determined and therefore the gauge coupling constant, unification
scale and gravitino mass are not free parameters.
 In practice, there are a large number
of consistent vacua having the standard model gauge group  and three
generations of particles and until  now there is has been no procedure to
choose one. To avoid this problem,
we work in a model independent form by
allowing the coefficients $n_{g}, b_{0}$ and $\alpha_{0}$ of the hidden sector
to be free
parameters from which the  the gauge coupling constant, unification
scale and gravitino mass are obtained. Again, it would seem that there
is not any    predictive power, since  we are replacing the three parameters
of the MSSM with  three other ones. Yet, this point of view is
not entirely fair since the new parameters can take only a limited
number of discrete values and in principle
they are not free at all, as argued above.
One could  further restrict the minimal string
unification by demanding that the gauge coupling constants of the
visible sector and of the hidden sector
become unified at the unification scale. In what follows we will
consider both possibilities; (i) that all gauge coupling constants are unified
at the same scale and (ii) that only the gauge coupling constants of
the SM are unified.
Let us consider  the (i) case first. From eq.(\ref{f21}), we notice that
if all gauge coupling constants are to be unified, then the $\alpha$
parameters of the hidden and visible sector must be the same, i.e.
$\alpha_{0}^{i}=\alpha_{a}^{i}=\alpha_{b}^{i}$. From the
minimization condition  we
obtained that
$\alpha_{0}^{i}=\alpha_{0}^{j}$ for those moduli with  "large" v.e.v.'s
and thus all $\alpha$'s must have the same value\footnote{For  moduli with
dual invariant v.e.v. the modular invariant quantity $
(T_{rk}|\eta(T_{k})|^{4})^{\alpha^{k}_{0}}\simeq 1$ and we will neglect it in
the unification scale.}.

Assuming that the gauge group in the hidden sector is of rank less or
equal eight (i.e. a subgroup of $E_{8}$) then $n_{g}\le 248$ (the
dimension of $E_{8}$), $0<16\pi^{2}b_{0}<90$ and $0<\alpha_{0}<1/3$. If the
hidden sector is broken down to an $SU(N)$ subgroup (with $N\le 9$)
then the coefficients are restricted to take values in the range
\bea
0&<16\pi^{2}\,b_{0}\le& 3N,
\la{ff24}\\
0&< \alpha_{0}<&1/3
\la{f24}\eea
and
\be
n_{g}=N^{2}-1.
\la{f25}\ee
Thus we see that a very small set of
values for $n_{g}, b_{0}$ and $\alpha_{0}$ are allowed and that the values
of the MSSM parameters (cf. eq.(\ref{f2}-\ref{f3})) that can be obtained is far
from trivial.
Furthermore, the structure constant at the unification scale is given
by
\be
\hat{\alpha}^{-1}_{gut}=\frac{16\pi^{2}}{\sqrt{n_{g}}}.
\la{ff1}\ee
{}From eqs.(\ref{f2}) and (\ref{ff1}),
a coupling constant consistent with the range  of MSSM  requires a
gauge group with $33.2<n_{g}<43.4$, which fixes the hidden gauge group
to be $SU(6)$ or $SO(9)$. For these gauge groups, the fine
structure constant would be $\hat{\alpha}^{-1}_{gut}\simeq 26.7$ and $26.3$
respectively.

To gain a better insight into the solution
 obtained in eqs(\ref{e30}-\ref{e31}) we
will trade the one-loop beta function coefficient of the hidden gauge
group for the gravitino
mass and we will rewrite the unification scale in terms of $m_{3/2}, n_{g}$
and $\alpha_{0}$. From eqs.(\ref{f21}) and (\ref{ff3}) we have
\bea
\Lambda_{gut}&=&M_{s}\Pi_{i}(T_{ri}|\eta_{i}|^{4})^{\alpha_{0}/2}
\nonumber\\
\Lambda_{gut}&=&M_{s}\left(\frac{4Ym_{3/2}^{2}}{M_{s}^{6}}
\right)^{\frac{\alpha_{0}}{2(3\alpha_{0}-1)}}\,\,e^{-\frac{3Y}{4b_{0}}\frac{1}{(1-3\alpha_{0})}}
\la{f26}\eea
and the $b_{0}$ dependence is in the exponent only. Using eq.(\ref{f16}),
one can eliminate this dependence and we get
\be
\Lambda_{gut}=M_{s}\lr[\frac{\xi^{3}B_{0}^{3}m_{3/2}}{32\,Y\,x^{3/2}_{g}}\rr]^{\alpha_{0}/2}
\la{f27}\ee
with $b_{0}$ given by
\be
b_{0}=-\frac{3Y}{2}\lr[\,ln\lr(4Y\frac{m_{3/2}^{2}}{M_{s}^{6}}\lr[\frac{\xi^{3}B_{0}^{3}m_{3/2}}{32Yx_{g}^{3/2}}\rr]^{1-3\alpha_{0}}\rr)\,\rr]^{-1}.
\la{f28}\ee
The r.h.s. of eqs.(\ref{f27}) and (\ref{f28}) still depend on $b_{0}$
through $x_{g},\xi$ and $B_{0}$ but the
unification scale depends on $b_{0}$ now only linearly.
It is not possible to   solve  eq.(\ref{f28})  analytically
for  $b_{0}$, although its dependence on the r.h.s. is  only
logarithmic. By setting  $16\pi ^{2}b_{0}=3N$ on the r.h.s.,
one obtains a  good approximation.

 We can
now plot $\Lambda_{gut}\,vs.\,\alpha_{0}$ for a fixed value of the gravitino
mass.
In fig.6.1 we show the graph
for an $SU(6)$ gauge group with $m_{3/2}=82 \, GeV$.
The unification scale has a minimum at around
$\alpha_{min}=0.3$  with a value of
\be
\Lambda_{gut}\simeq 2.8\times 10^{16}\,GeV.
\la{f29}\ee
It is remarkable that the minimum value for the unification scale is just about
the
value required  by the minimal unification of couplings in the
MSSM. The value of $\alpha_{min}$ does not depend on the gravitino
mass. For decreasing  value of the gravitino mass, the unification scale is
also  reduced as it is clear from eq.(\ref{f27}). If we require that
$m_{3/2}>35\,GeV$, then $16\pi^{2}\,b_{0}>14.8$.  Since $b_{0}$
 can only  take
discrete values, we have set  $16\pi^{2}\,b_{0}=15$ which
gives $m_{3/2}\simeq 82\, GeV$.
The value of the moduli for this specific example are $T_{r}/2=22.3,12.1,8.7$
for one, two or three moduli with "large" v.e.v., respectively.
 It  is precisely the fact that the v.e.v. of the moduli
get a ``large'' v.e.v. that permits the unification scale to be much
smaller then the string scale (note that for
$T_{r}\simeq 1.2$ as obtained at tree level the unification scale will
always be of the same order of magnitude as the string scale).
Furthermore, these
solutions allow for squeezed orbifolds which where found
better candidates for minimal string unification.

We will now consider the case in which the hidden sector gauge coupling
constant does not necessarily become unified with the gauge coupling
constants of the standard model. This would be the case when the threshold
corrections to the gauge coupling constant differ for the hidden and visible
 sector gauge groups.
In this case, there is no connection
between the $\alpha$ parameters of the hidden sector and the ones in
the visible sector. Thus, the unification scale is
\be
\Lambda_{gut}=M_{s}\lr[\frac{\xi^{3}B_{0}^{3}m_{3/2}}{32\,Y\,x^{3/2}_{g}}\rr]^{\bar{\alpha}/2}.
\la{f31}\ee
The term in square brackets  is independent of $\alpha_{a}$ and
$\bar{\alpha}$ is  the average of the $\alpha_{a}^{i}$ whose
corresponding moduli $T_{i}$ get a ``large'' v.e.v., i.e.
$\bar{\alpha}\equiv
\frac{1}{n_{l}}\Sigma_{i}^{n_{l}}\alpha_{a}^{i}=\frac{1}{n_{l}}\Sigma_{i}^{n_{l}}\alpha_{b}^{i}$,
where $n_{l}$ is the number of moduli with ``large'' v.e.v.
It is easy to see that in this case a unification scale consistent
with MSSM is possible. The value of $\alpha_{0}$ is restricted  only to
be smaller  then 1/3 and it can take negative values. As an example, we
could consider a hidden gauge group with matter field in the untwisted
sector only and taking $\delta_{GS}=0$ one has $\alpha_{0}=-1/3$. For
this value of $\alpha_{0}$ and choosing $\bar{\alpha}=0.32$, one obtains
a gravitino mass  and a unification scale of
\bea
m_{3/2}&\simeq & 632\,GeV
\\
\Lambda_{gut}&\simeq & 2 \times 10^{16}\,GeV.
\la{f32}\eea
In this example the v.e.v of the moduli are $T_{r}/2=ReT=22,\,12.3,\,9$
for one, two or three with ``large'' v.e.v. and $16\pi^{2}b_{0}=11$.
The same value  of $\Lambda_{gut}$ can be obtained for larger
gravitino mass and  $\bar{\alpha}$.

To conclude, we have shown  that after finding a stable solution
to the scalar potential
including the contribution from gaugino binding effects, the v.e.v.
obtained for the moduli and dilaton fields
give the required values of MSSM for
the gauge coupling constant  and supersymmetry breaking scale (i.e.
gravitino mass). The unification scale can also be consistent with
MSSM for specific values of $\alpha$.
Furthermore, since the value of the fine structure constant at the
unification scale is fixed by the dimension of the hidden sector gauge group
this group  must be $SU(6)$ or $SO(9)$.


{99}

\newpage

\begin{figure}
\vspace{12cm}
\caption{Unification scale as a function of $\alpha$, keeping $m_{3/2}$
fixed.}
\end{figure}

\end{document}